\documentclass[aps, longbibliography,floatfix,amsmath,amssymb,amsfonts,notitlepage, twocolumn,superscriptaddress,prr]{revtex4-2} 
\usepackage[T1]{fontenc}
\usepackage{lmodern} 
\usepackage{xcolor}
\usepackage{bbold}
\usepackage{dsfont}
\usepackage{graphicx}
\usepackage[caption=false]{subfig}
\usepackage[colorlinks]{hyperref}
\hypersetup{colorlinks,linkcolor={blue},citecolor={blue},urlcolor={red}}
\usepackage[bottom]{footmisc}
\usepackage{enumerate}
\usepackage{float}
\usepackage{mathtools}
\usepackage{bm}
\usepackage[draft,inline,nomargin]{fixme}
\usepackage[normalem]{ulem}
\usepackage{balance}

\usepackage{diagbox}

\DeclarePairedDelimiter\bra{\langle}{\rvert}
\DeclarePairedDelimiter\ket{\lvert}{\rangle}

\DeclarePairedDelimiterX\braket[2]{\langle}{\rangle}{#1 \delimsize\vert #2}
\DeclarePairedDelimiterX\ketbra[2]{\vert}{\lvert}{#1 \delimsize\rangle\langle #2}

\newcommand{\sg}[2]{ \hat{\sigma}_{#1}^{(#2)}}
\newcommand{\ee}[1]{ e^{#1}}

\begin{document}
\newcommand{\todo}[1]{\red{{$\bigstar$\sc #1}$\bigstar$}}
\newcommand{\ks}[1]{{\textcolor{teal}{[KS: #1]}}}
\newcommand{\pbb}[1]{{\textcolor{blue}{[pbb: #1]}}}
\global\long\def\eqn#1{\begin{align}#1\end{align}}
\global\long\def\ket#1{\left|#1\right\rangle }
\global\long\def\bra#1{\left\langle #1\right|}
\global\long\def\bkt#1{\left(#1\right)}
\global\long\def\sbkt#1{\left[#1\right]}
\global\long\def\cbkt#1{\left\{#1\right\}}
\global\long\def\abs#1{\left\vert#1\right\vert}
\global\long\def\der#1#2{\frac{{d}#1}{{d}#2}}
\global\long\def\pard#1#2{\frac{{\partial}#1}{{\partial}#2}}
\global\long\def\re{\mathrm{Re}}
\global\long\def\im{\mathrm{Im}}
\global\long\def\dd{\mathrm{d}}
\global\long\def\ddd{\mathcal{D}}

\global\long\def\avg#1{\left\langle #1 \right\rangle}
\global\long\def\mr#1{\mathrm{#1}}
\global\long\def\mb#1{{\mathbf #1}}
\global\long\def\mc#1{\mathcal{#1}}
\global\long\def\tr{\mathrm{Tr}}
\global\long\def\dbar#1{\Bar{\Bar{#1}}}

\global\long\def\nth{$n^{\mathrm{th}}$\,}
\global\long\def\mth{$m^{\mathrm{th}}$\,}
\global\long\def\non{\nonumber}
\newcommand{\teal}[1]{{\color{teal} {#1}}}

\newcommand{\orange}[1]{{\color{orange} {#1}}}
\newcommand{\cyan}[1]{{\color{cyan} {#1}}}
\newcommand{\blue}[1]{{\color{blue} {#1}}}
\newcommand{\yellow}[1]{{\color{yellow} {#1}}}
\newcommand{\green}[1]{{\color{green} {#1}}}
\newcommand{\red}[1]{{\color{red} {#1}}}
\global\long\def\todo#1{\yellow{{$\bigstar$ \orange{\bf\sc #1}}$\bigstar$} }

\title{Delay-induced spontaneous dark- state generation from two distant excited atoms}

\author{W. Alvarez-Giron}
\email{wikkilicht@ciencias.unam.mx }
\affiliation{Instituto de Investigaciones en Matem\'{a}ticas Aplicadas y en Sistemas, Universidad Nacional Aut\'{o}noma de M\'{e}xico, Ciudad Universitaria, 04510, DF, M\'{e}xico.}

\author{P. Solano}
\email{psolano@udec.cl}
\affiliation{Departamento de F\'isica, Facultad de Ciencias F\'isicas y Matem\'aticas, Universidad de Concepci\'on, Concepci\'on, Chile}
\affiliation{CIFAR Azrieli Global Scholars program, CIFAR, Toronto, Canada.}

\author{K. Sinha}
\email{kanu@arizona.edu}

\affiliation{College of Optical Sciences and Department of Physics, University of Arizona, Tucson, Arizona 85721, USA}

\author{P. Barberis-Blostein}
\email{pablobb@gmail.com }
\affiliation{Instituto de Investigaciones en Matem\'{a}ticas Aplicadas y en Sistemas, Universidad Nacional Aut\'{o}noma de M\'{e}xico, Ciudad Universitaria, 04510, DF, M\'{e}xico.}

\begin{abstract}
  We investigate the collective non-Markovian dynamics of two
  fully excited two-level atoms coupled to a one-dimensional waveguide
  in the presence of delay. We demonstrate that analogous to the
  well-known superfluorescence phenomena, where an inverted atomic
  ensemble synchronizes to enhance its emission, there is a
  ``subfluorescence'' effect that synchronizes the atoms into an
  entangled dark state depending on the interatomic separation.
  The phenomenon can lead to a two-photon bound state in the
    continuum. Our results are pertinent to long-distance quantum
  networks, presenting a mechanism for spontaneous entanglement
  generation between distant quantum emitters.
\end{abstract}

                 
\maketitle

\section{Introduction}
Large-scale interconnected quantum systems offer promising
applications in quantum information processing and distributed quantum
sensing~\cite{quantum-network-1, quantum-network-2,
  quantum-network-3}. As the experimental capabilities for directly
interconnecting distant quantum nodes grow \cite{Storz2023}, we have
yet to develop the theoretical toolbox to efficiently analyze systems
of multiple qubits collectively exchanging photons over large
distances, where delay effects cannot be neglected. Such many-body
systems exhibit rich non-Markovian and nonlinear dynamics, arising
from delayed and multiphoton interactions. Given the complexity
associated with such systems, even the simplest scenarios, e.g., the
spontaneous decay of two fully excited distant atoms, remains
unexplored.

The most studied cases of coupled emitters consider nearby atoms,
neglecting the delay time the field needs to propagate between them.
This scenario is a landmark of quantum optics across platforms, such
as atoms in free space \cite{superradianceessay,PhysRevA.68.023809},
inside a leaky cavity \cite{carmichael-1}, and near waveguides
\cite{Lekienmodes}. In all of these cases, ensembles of initially
fully excited two-level atoms decay into their ground state. As atoms
decay, correlations among them spontaneously and momentarily emerge,
leading to the well-known collective phenomenon of
\textit{superfluorescence} ~\cite{Bonifacio75, Glauber78,
  vrehen1980,PhysRevA.68.023809}\footnote{We use the term
  superfluorescence~\cite{Bonifacio75} as a particular case of
  \textit{superradiance} when considering a fully inverted sample.
  Broadly speaking, \textit{superradiance} is the collective
  enhancement of the decay rate of an atom in an atomic ensemble
  beyond the spontaneous emission rate of a single
  atom~\cite{Dicke,superradianceessay}}. Such atom-atom correlations,
which are absent in the initial state, emerge without external driving
fields or post-selection ~\cite{superradianceessay}.

Platforms based on waveguide QED present an effective way to go beyond
the zero-delay approximation, enabling tunable, efficient, and
long-ranged interactions in the optical and microwave regime
\cite{Goban15,Solano2017,PabloReview,Kim2018,Wen19,Mirhosseini19,Sheremet2021,Magnard2022,Tiranov2023}.
For example, state-of-the-art experiments allow one to tune out dispersive dipole-dipole interactions by strategically positioning the atoms along a waveguide \cite{Martin-Cano11, VanLoo13, Pichler15} while highly reducing coupling to non-guided modes \cite{Lecamp07,astafiev2010, Arcari14, Zang16, Scarpelli19}. Furthermore, waveguide-coupled quantum emitters enable chiral interactions due to strong spin-momentum coupling, and  directional routing of photons~\cite{Lodahl2017,Kannan2023, Solano2023, maffei2024,Lin19,Petersen2014}.

When the time taken by a photon to propagate between two distant emitters becomes comparable to their characteristic lifetimes, the system exhibits surprisingly rich delay-induced non-Markovian dynamics even in the single-photon regime \cite{Zheng13, Tufarelli14, Guimond16, Zhang20,DelAngel22,Solano2023}. Some examples of such dynamics include collective spontaneous emission rates exceeding those of Dicke superradiance and formation of highly delocalized atom-photon bound states \cite{sinha, Sinha20, Dinc19,   Calajo19, Facchi19, Guo19, Hughes09, Hughes2020, Lee23, Carmele2020}. In addition, time-delayed feedback can assist in preparing photonic cluster states \cite{Pichler17} and single-photon sources with improved coherence and indistinguishability \cite{Crowder23}.
  
\begin{figure}[b]
\centering
\includegraphics[width=0.45\textwidth]{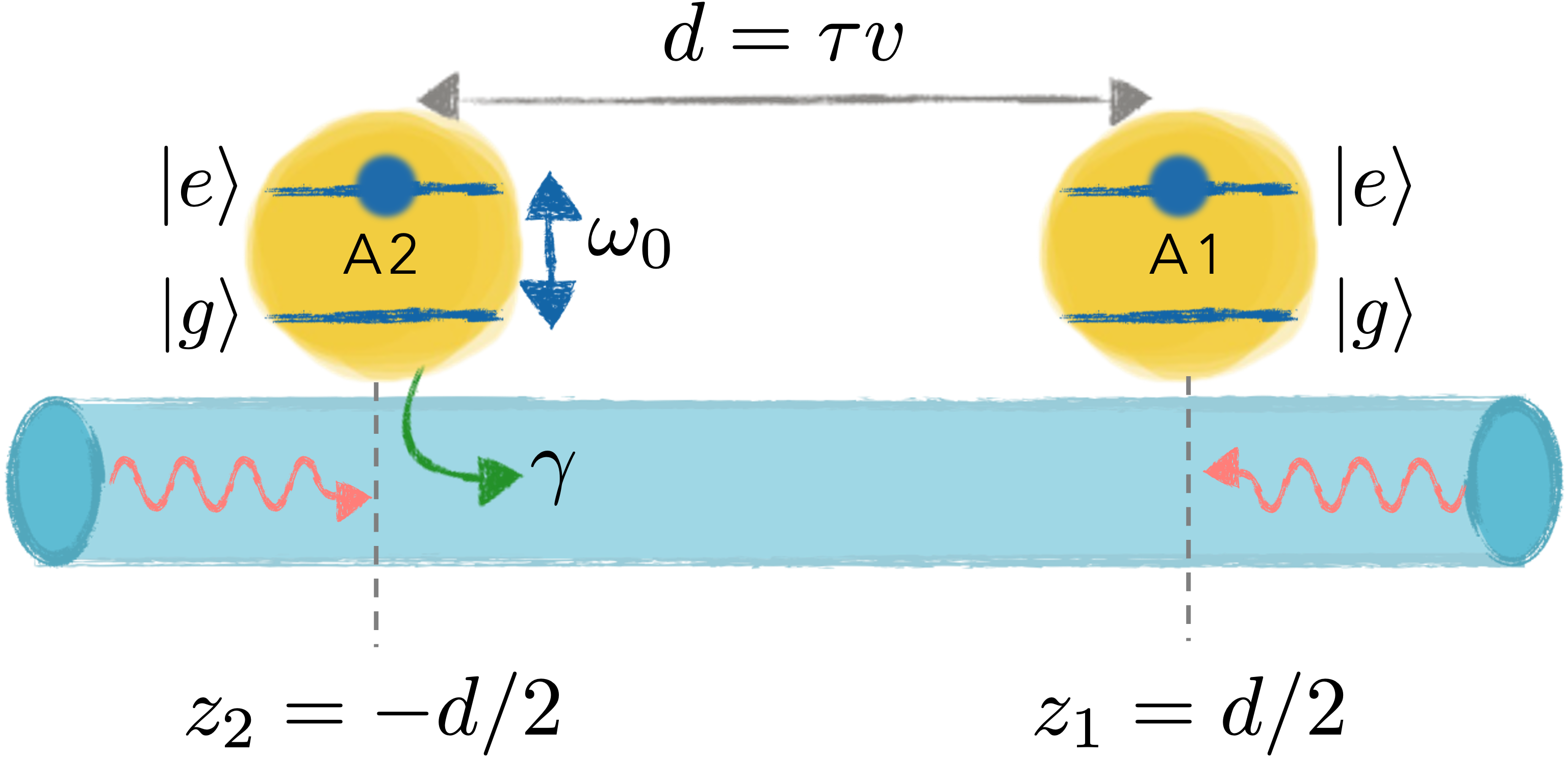}
\caption{Schematic of the system: two excited atoms with a transition
  frequency $\omega_0$ near a waveguide at positions   $z_{1,2} = \pm
  d/2$. The emission rate of each atom into   the waveguide is
  $\gamma$ and the propagation delay of the field between the two atoms is $\tau = d/v$, with $v$ as the velocity of
  the EM field in the waveguide.}
\label{fig:system}
\end{figure}

Recent studies of the effects emerging from delayed interactions considered either many atoms with a single photon \cite{Gonzalez2013, Liao15, Facchi19, sinha, Pivovarov21}, many photons with a single atom \cite{Calajo19, Guo17, Guimond17, Fang18}, or numerical exploration of three-photons and three-atoms \cite{Carmele2020}, inviting us to revisit the theoretical description of other canonical quantum optical phenomena, such as superfluorescence, in the context of waveguide-coupled emitters in a large-scale quantum network. 

In this paper, we show that a system of two-atoms, with
two-excitations, and delayed-interactions (see Fig.~\ref{fig:system})
exhibits novel non-Markovian collective dynamics. In particular, for
fully inverted atoms, the instantaneous decay can be faster than in
superfluorescence and steady-state entanglement can suddenly emerge.
In the latter case, the electromagnetic (EM) field can have two
photons trapped between the two atoms, an effect that can not be
captured by Markovian dynamics.
We refer to this phenomenon, which leads to the spontaneous generation
of a dark state, as \textit{delayed-induced subfluorescence} --the
subradiant counterpart of the well-known superfluorescence effect. The
resulting steady state is a delocalized hybrid atom-photon state,
requiring a description beyond the usual Born and Markov
approximations and qualitatively different from the case without
delay, where the system always ends in the ground state.
The rich and unique dynamics in such a seemingly simple system demonstrates non-Markovian delay as a mechanism for trapping light, entanglement generation and stabilization in long-distance
interacting quantum nodes.

\section{Theoretical model}
We consider two two-level atoms, with transition frequency $\omega_0$ between the ground $\ket{g}$ and excited $\ket{e}$ states, coupled through a waveguide. In this work, our emphasis is on the case of perfectly coupled atoms, the free Hamiltonian of the total atoms+field system is given by $H_0=(\hbar\omega_0/2)\sum_{r=1}^2\sg{z}{r}+\sum_{\alpha, \eta}\hbar\omega_\alpha\hat{a}^\dagger_{\alpha,   \eta}\hat{a}_{\alpha, \eta}$. The interaction Hamiltonian in the interaction picture is (having made the electric-dipole and rotating-wave approximations):
\begin{eqnarray}
\hat{H}_\text{int} = -i \hbar \sum_{r=1}^2 \sum_{\alpha, \eta}
 g_\alpha \sg{+}{r} \hat{a}_{\alpha, \eta} \ee{-i\Delta_\alpha t} \ee{ik_{\alpha\eta} z_r} 
\label{ham} 
+ \text{H.c.} \, ,
\end{eqnarray}
where $g_\alpha$ is the coupling between atoms and the guided mode $\alpha$ with frequency $ \omega_\alpha$, $\Delta_\alpha = \omega_\alpha - \omega_0$, and $k_{\alpha \eta}= \eta\omega_\alpha/v$, with $\eta = \pm 1$ specifying the direction of propagation. The position of the atoms is $z_{1, \,2} = \pm d/2$, $\sg{z}{r}=\ket{e}_r\bra{e}_r-\ket{g}_r\bra{g}_r$ and the raising and lowering operators for the $r$-th atom are defined as $\sg{+}{r} = \bkt{\sg{-}{r}}^\dagger = \ket{e}_r\bra{g}_r$.

We consider the initial state with both atoms excited and the field in the vacuum state, $\ket{e_1 e_2,\{0\}}$. As a consequence of the rotating-wave approximation, the total number of atomic and field excitations are conserved, suggesting the following ans\"atz for the state of the system at  time $t$:
\begin{widetext}
    
  \begin{align}
    \ket{\psi (t)} = \bigg\{ a(t) \sg{+}{1}\sg{+}{2} + \sum_{r=1}^2 \sum_{\alpha, \eta}b_{\alpha \eta}^{(r)}(t) \sg{+}{r} \hat{a}_{\alpha, \eta}^\dag + \mathop{\sum \sum}_{\alpha , \eta\neq \beta, \eta'} \frac{c_{\alpha\eta, \beta  \eta'}(t)}{2} \hat{a}_{\alpha, \eta}^\dag \hat{a}_{\beta, \eta'} ^\dag
+ 
\sum_{\alpha, \eta} \frac{c_{\alpha \eta}(t)}{\sqrt{2}} \hat{a}_{\alpha, \eta}^\dag \hat{a}_{\alpha, \eta} ^\dag
\bigg\} \ket{g_1 g_2,\{0\}} \, ,
\label{eq:psi}
\end{align}
\end{widetext}
where $\ket{g_1g_2,\{ 0\}}$ is the ground state of the system. The complex coefficients $a(t)$, $b_{\alpha \eta}^{(r)}(t)$, and $c_{\alpha\eta, \beta\eta'}(t)$ correspond to the probability amplitudes of having an excitation in both the atoms, $r^\mr{th}$ atom and field mode $\cbkt{\alpha, \eta}$, and field modes $\cbkt{\alpha, \eta}$ and $\cbkt{\beta, \eta'}$, respectively. The coefficients $c_{\alpha\eta}(t)$ represent the probability amplitude of exciting two photons in modes $\cbkt{\alpha, \eta}$. The Hamiltonian in Eq.~\eqref{ham} models two emitters interacting via the quantized electromagnetic field; their distance, codified in the phase $\ee{ik_{\alpha\eta} z_r}$, has the effect of introducing a delay in the equations of motion for the quantum state coefficients. This delayed interaction between the emitters results from the finite propagation speed of light.

Defining $B_{\alpha \eta}^{(r)} = b_{\alpha \eta} ^{(r)} (t) \ee{-i\Delta_\alpha t}$ and $C_{\alpha\eta} = c_{\alpha\eta} (t) \ee{-2i\Delta_\alpha t}$, and formally integrating the equation for $c_{\alpha \eta,\beta \eta'}(t)$, the Schr\"odinger equation yields the following system of delay-differential equations for the
excitation amplitudes: \eqn{
\label{a-2}
\dot{a}(t) =& - \sum_{\alpha,\eta} \sum_{s=1}^2 g_\alpha B_{\alpha\eta}^{(s)}(t) \ee{-ik_{\alpha\eta} z_s}\, ,
}
\eqn{
\dot{B}_{\alpha\eta}^{(r)}(t) =& -\bkt{i\Delta_\alpha + \frac{\gamma}{2}} B_{\alpha \eta}^{(r)}(t) 
-\sqrt{2} g_\alpha 
C_{\alpha\eta}(t) \ee{ik_{\alpha\eta} z_r} +
\nonumber \\
&  
 g_\alpha^\ast a(t)
\ee{ik_{\alpha\eta} z_r}  - \frac{\gamma}{2} \ee{i\phi} \ee{-i\Delta_\alpha t} B_{\alpha\eta} ^{(s)}(t-\tau)  \Theta(t-\tau)
\label{b-2}\\
\label{c1-2}
\dot{C}_{\alpha\eta} (t) =& - 2i\Delta_\alpha C_{\alpha\eta}(t)  + \sqrt{2} g_\alpha ^\ast\sum_{s=1}^2 B_{\alpha \eta}^{(s)}(t) \ee{-ik_{\alpha\eta} z_s} \, ,
}
where  $\gamma$ is the decay rate of the emitters into the guided modes. We present a detailed derivation in Appendix \ref{app1}. We solve the system of equations via Laplace transforms considering the initial conditions $a(0) = 1$ and $B_{\alpha\eta}(0)=C_{\alpha\eta}(0)=0$.

\section{System dynamics} The Laplace transform of $a(t)$ is (see Appendix~\ref{appsolution})
\begin{widetext}
\begin{eqnarray}
\label{a12-solapp}
\tilde{a}(s) 
&\approx & \left\{s - \sum_{\alpha, \eta} |g_\alpha |^2  \frac{s + i\Delta_\alpha + \frac{\gamma}{2} - \frac{\gamma}{2}e^{i\phi} e^{-s \tau }e^{ - i \Delta_\alpha \tau} \cos(\omega_\alpha \tau)}{\left[  \frac{\gamma}{2}e^{i\phi} e^{-s \tau }e^{ - i \Delta_\alpha \tau} + s + i\Delta_\alpha + \frac{\gamma}{2} \right] \left[  \frac{\gamma}{2}e^{i\phi} e^{-s \tau }e^{ - i \Delta_\alpha \tau} - s - i\Delta_\alpha - \frac{\gamma}{2}
\right] } \right\}^{-1} \, ,
\end{eqnarray}
\end{widetext}
where $\phi = \omega_0 \tau$ is the resonance field propagation phase.

\begin{figure}[b]
\centering
\includegraphics[width=0.45\textwidth]{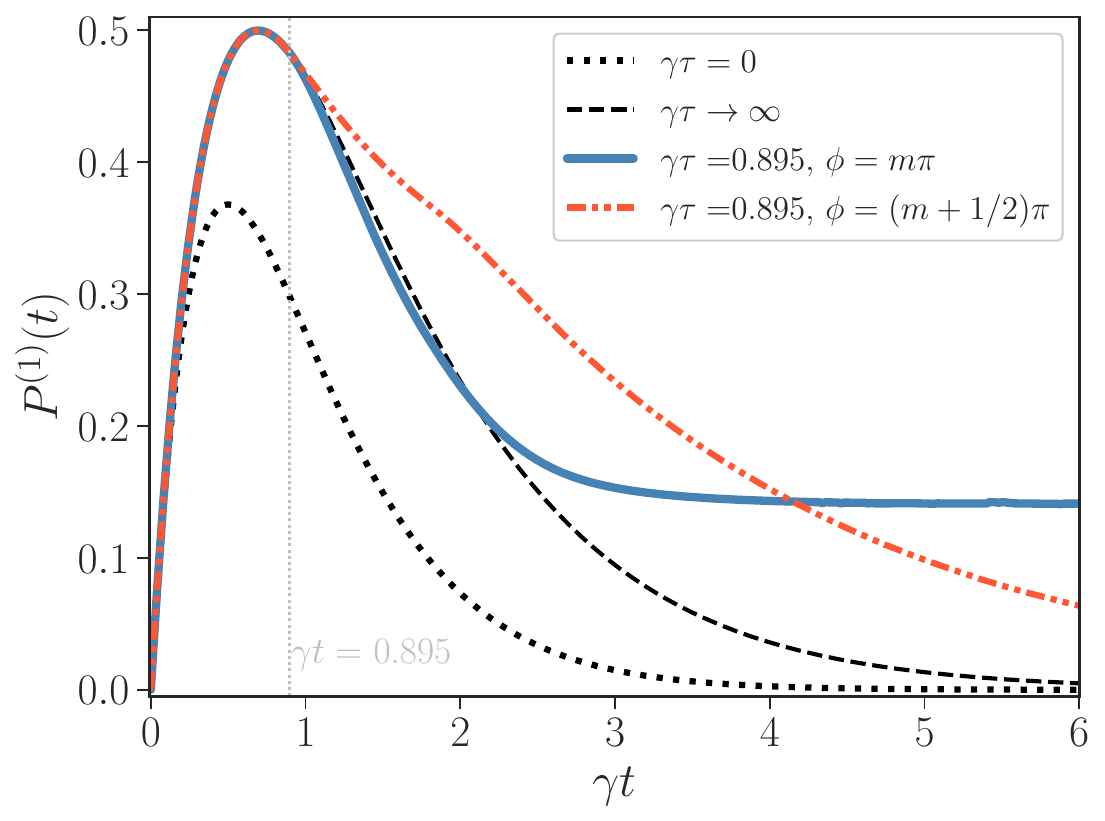}
  \caption{\label{fig:P1_dynamics} Probability $P^{(1)}(t)$ of having only one of the atoms excited for a delay of $ \gamma\tau = 0.895$ compared to that of coincident and independent atoms ($\gamma \tau = 0$ and $\gamma \tau \to\infty $, respectively). We note that for a propagation phase of $ \phi = n\pi $, there is a formation of a steady atom-photon bound state as indicated in Eq.~\eqref{eq:boundstate}. }
\end{figure}

We write the sum over modes on the right-hand side (RHS) of Eq.
\eqref{a12-solapp} as an integral over frequencies by introducing the
density of modes $\rho(\omega_\alpha)$. Additionally, we use the
Wigner-Weisskopf approximation and set
$\rho(\omega_\alpha), \, g_\alpha \approx \rho(\omega_0), \, g_0$,
evaluated at the atomic resonance frequency, a good approximation in
the presence of delay effects in waveguide QED \cite{DelAngel22}. With
these considerations we obtain $\tilde{a}(s) = 1/(s+\gamma)$ for all
$\tau$. Taking the inverse Laplace transform we get
$a(t)= e^{-\gamma t}$, which gives the time-dependent probability of
having two atomic excitations
\begin{eqnarray}
P^{(2)}(t) = \left\lvert a(t) \right\rvert^2 = e^{-2\gamma t} \, .
\end{eqnarray}
We remark  that the probability of having two atomic excitations is independent of the delay between atoms.

\label{sec-b}

The probability of having only one of the atoms excited is

\begin{eqnarray}
P^{(1)}(t) \approx  \rho(\omega_0) \sum_{r=1,2} \sum_{\eta }\int_{0}^{\infty} d\omega_\alpha \,
\left \lvert b_{\alpha\eta}^{(r)}(t)\right \rvert ^2 
  ,
\label{p1-sol}
\end{eqnarray}
where the time-dependent solutions $b_{\alpha\eta}^{(1,2)}(t)$ are derived in Appendix \ref{appsolution} and given by (\ref{bt}). In the limits of  two coincident atoms with $\tau= 0$, and two infinitely distant atoms with  $\tau \to \infty$, we obtain the expected Markovian dynamics \cite{Lehmberg1970, alvarez}:
\begin{equation}
\label{p1-cases}
P^{(1)}(t) = \begin{cases}
2 \gamma t \,\ee{-2\gamma t} &\text{for  $\tau = 0$}\\
2 \ee{-\gamma t} (1 - \ee{-\gamma t}) &\text{for $\tau \to \infty$}
\end{cases} \, . 
\end{equation}

Figure~\ref{fig:P1_dynamics} shows $P^{(1)}(t)$ for different values of the delay $\tau$ and the propagation phase $\phi$. For times $t < \tau$, the atoms decay independently. At time $t = \tau$, the field emitted by one atom reaches the other, modifying their decay dynamics depending on the value of the propagation phase $\phi$. Their instantaneous decay rate, given by the negative slope of $P^{(1)}(t)$, could momentarily exceed that of standard superfluorescence ($\tau=0$). The delay condition $\gamma\tau \approx 0.375$ and $\phi=n\pi$ maximizes the instantaneous decay rate. This phenomenon is a signature of \textit{superduperradiance}, reported in Ref.~\cite{sinha}. Notably, in this case the atom-atom coherence that is necessary to modify the instantaneous decay 
emerges spontaneously.

In the late-time limit, the following steady state appears when $\phi = n\pi$ for any delay $\tau$ between the atoms (see Appendix~\ref{appstationary}):
\eqn{
  \label{eq:boundstate}
  \ket{\psi (t\to \infty)} =\mathop{\sum \sum}_{\alpha, \eta \neq \beta, \eta'} \frac{c_{\alpha\eta, \beta\eta'}(t \to \infty)}{2}  \ket{g_1 g_2,\{1_{\alpha,\eta}1_{\beta,\eta '}\}} &\non\\
 +\ket{e_1 g_2}\sum_{\alpha,\eta} b_{\alpha\eta}^{(1)} \ket{1_{\alpha,\eta}}  +\ket{g_1 e_2}\sum_{\alpha, \eta} b_{\alpha\eta}^{(2)}
  \ket{1_{\alpha,\eta}}&\, ,
}
where we have removed the explicit time dependence of the steady-state coefficients. The steady-state amplitude for the mode $\cbkt{\alpha,\eta}$ is
\begin{equation}
b_{\alpha\eta}^{\bkt{\substack{1\\2}}}= \mp\eta\ee{\pm i \eta n \pi /2}\frac{g_\alpha^{*}}{ 1 + \frac{\gamma \tau}{2} }
\frac{  i\sin\left(  \Delta_\alpha \tau /2\right)}{\gamma - i
  \Delta_\alpha} \, ,
  \label{eq:boundstatecoef}
\end{equation}
and $c_{\alpha\eta, \beta\eta'}$ is given in Eq.~\eqref{Eq:cinf}in  Appendix~\ref{appstationary}.  The last two terms in Eq.~\eqref{eq:boundstate} represent a bound state in the continuum (BIC) that corresponds to having one shared excitation between the atoms and one propagating photon mode in between them \cite{Calajo19}. Using the Born rule and Eq.~(\ref{eq:boundstatecoef}), we obtain that its probability is $2 P^{(1)}\bkt{t\to\infty}$, where:
\begin{align}
\label{p1ss}
P^{(1)}\bkt{t\to\infty} = \frac{\sinh \left(\frac{\gamma\tau}{2}
  \right)}{(1+\frac{\gamma\tau}{2})^2 \ee{\frac{\gamma\tau}{2}}}\, .
  \end{align}

We note that the probability of ending up in a BIC state is maximum for $\gamma \tau \approx 0.895$ with a probability of $0.282$.

The fact that the atoms+field state is non-separable shows that one
cannot use the Born-Markov approximation to solve for the dynamics of the system. Using Eq.~(\ref{eq:boundstate}) and Eq.~(\ref{eq:boundstatecoef}) we obtain the following reduced density
matrix for the atomic subsystem
\begin{equation}
  \label{eq:reducedrho}
  \hat{\rho}_A^\pm(t\to\infty)=P^{(1)}\ket{\Psi^\pm}\bra{\Psi^\pm}+\bkt{1-P^{(1)}}\ket{g_1
    g_2}\bra{g_1 g_2}\, ,
\end{equation}
with $+$ corresponding to the case when $n$ is odd and $-$ when it is even and $\ket{\Psi^\pm}=(\ket{e_1 g_2}\pm\ket{g_1e_2})/\sqrt{2}$ are single excitation Bell states.  As the initially inverted atoms evolve into a superposition of radiative and  non-radiative states upon delayed-interactions, the atoms decay into a superposition of ground state and an entangled dark steady state. For a field propagation phase $\phi =2 n\pi$ $(\phi = (2n+1) \pi)$, the dark state that appears in the late-time limit corresponds to the antisymmetric state $\ket{\Psi^- }$ (symmetric state  $\ket{\Psi^+}$). 

\begin{figure}[t]
\centering
\subfloat[]{\includegraphics[width=0.42\textwidth]{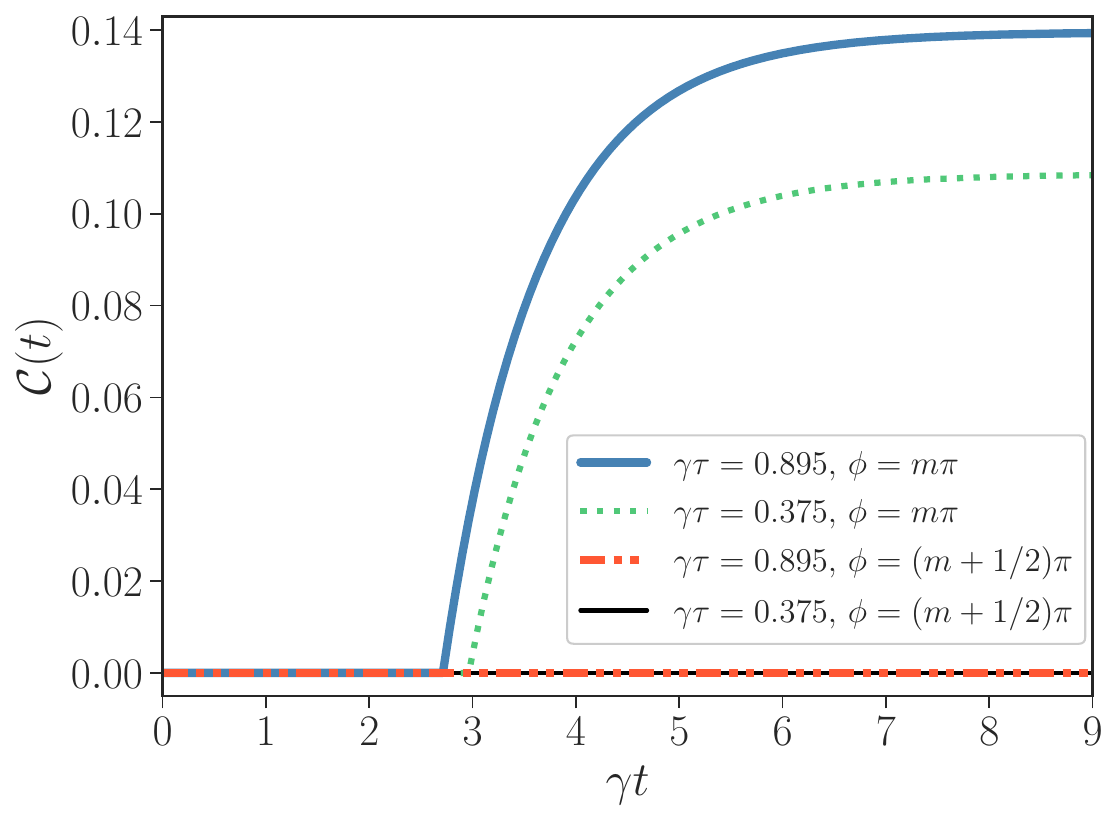}}\\
\subfloat[]{\includegraphics[width=0.45\textwidth]{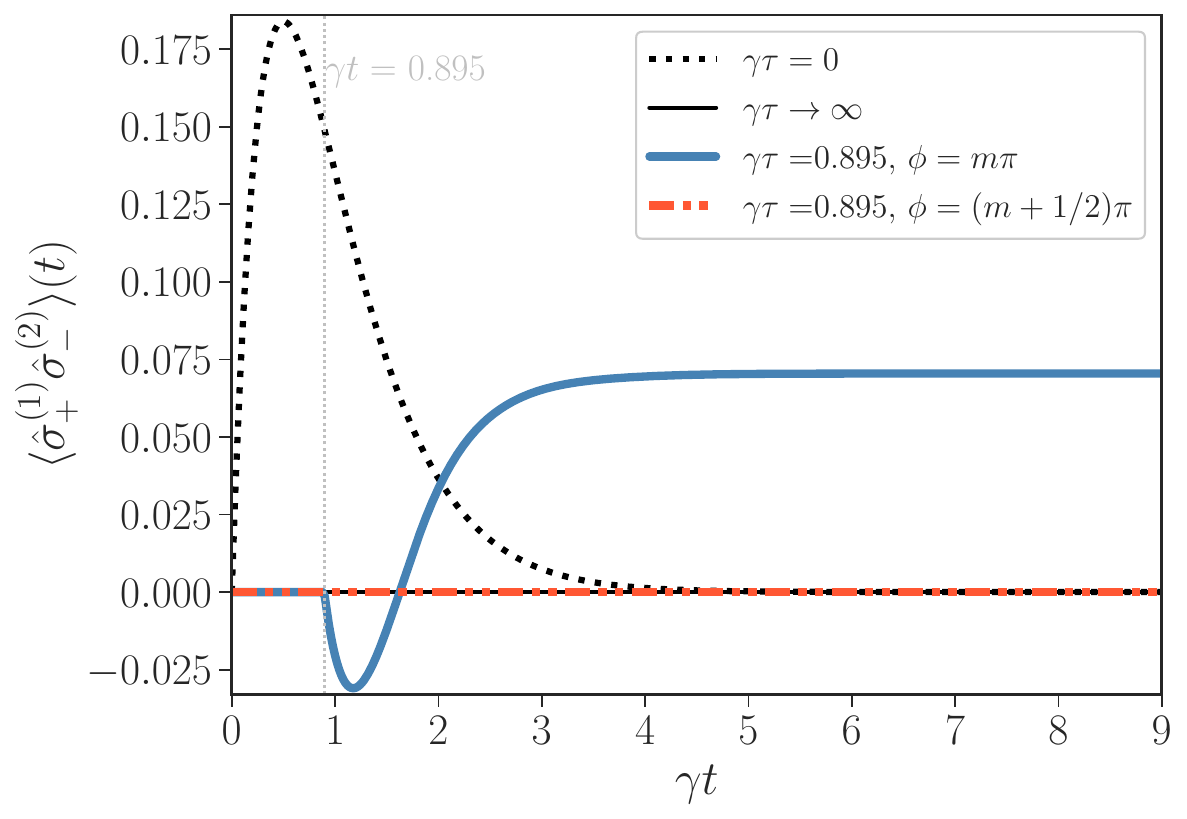}}
\caption{\label{fig:cmax}(a) Concurrence as a function of time. The atoms do not  have initial correlations $\mathcal{C}(0)=0$, after some time   $t_\mr{SBE}$ there is a sudden birth of entanglement”.  For $\phi = (m+1/2)\pi $ $\mathcal{C}(t) = 0$ for all time. If  $\phi = m\pi$ concurrence reaches a stationary value $\mathcal{C}_\text{ss}$, given by Eq.~\eqref{p1ss}. (b) Correlation  between atomic dipoles. The atoms decay independently for times  $t<\tau$. At $t=\tau$, we observe an emergence of quantum correlations between the  atoms. When $\phi=m\pi$, the atoms reach a dark steady state for all delays $\tau$.   The maximum  concurrence is $ \approx 0.141$ at $\gamma \tau \approx 0.895$.
 }
\end{figure}

Delayed interactions create quantum correlations between the atoms. We
quantify it using the concurrence $\mathcal{C}(t)$ of the reduced
density matrix of the atoms
$\hat{\rho}_A(t) = Tr_F(\ketbra{\psi(t)}{\psi(t)})$ \cite{Wooters}.
For $\tau \to \infty$, the concurrence is zero throughout the
evolution since the initially uncorrelated atoms evolve independently.
Remarkably, when $\tau = 0$, the concurrence is also zero throughout
the evolution, even when atoms transition to a superradiant behavior.
This case exemplifies that entanglement is not necessary for
superradiance. For intermediate values of $\tau$, we numerically study
the dynamics of concurrence, shown in Fig.~\ref{fig:cmax}~(a). We
begin studying its behavior for $\gamma \tau \approx 0.895$
(corresponding to the maximum value of $P^{(1)}\bkt{t\to\infty}$) and
$\gamma \tau \approx 0.375$ (corresponding to the largest
instantaneous decay rate). Since the system lacks initial
correlations, the concurrence is zero from $t=0$ to a certain time,
$t_\text{SBE}$, when there is a sudden birth of entanglement (SBE)
\cite{SBE1, SBE2, SBE3, SBE4,SBE5,SBE6,SBE7}. For $\phi = n\pi$, the
concurrence increases until reaching a stationary value, whereas for
$\phi =(n+\frac{1}{2}) \pi$, it always remains zero. In general, for
other values of $\phi$, the concurrence suddenly departs from zero and
slowly decays after reaching a maximum value. For
$\gamma \tau \approx 0.895$, we obtain the maximum value for the
concurrence. Figure~\ref{fig:cmax}~(b) shows the emergence of atom-atom
correlations defined by
$\tr\sbkt{\hat \rho_A (t)\hat{\sigma}_+ ^{(1)} \hat \sigma_-^{(2)} }$.
We note that the atom-atom correlations develop as soon as the atoms
``see'' each other, while the concurrence takes longer to emerge.
Delayed atom-atom interactions couple the fully excited state of the
atoms to both single excitation symmetric and antisymmetric states.
After the build-up of correlations, the most radiative state
collectively decays, while the non-radiative atomic state remains. The
system, therefore, spontaneously evolves into an entangled dark state
in the presence of retardation. Such appearance of quantum
correlations is a striking example of environment-assisted spontaneous
entanglement generation.

\section{Field Dynamics}
We now focus on the dynamics of the field sourced by the non-Markovian
behavior of the atoms. Field operators are often described through
mode expansion, as shown in Eq.~(\ref{fieldop}), or Green's functions
\cite{PhysRevA.53.1818}. Its expectation values involve integrating
over all the modes or knowing the Green's function for the field, which
for waveguides is usually given by a mode expansion \cite{Asenjo2017}.
When the system is Markovian, an input-output relation simplifies the
calculation of field observables, and the sum over field modes is
replaced by a sum over atomic operators:
$\hat{\mb E}^+(\mb r)= {\hat{\mb E}}^+_\textrm{Input}(\mb{r})+\sum_i
f(\mb {r},\mb {r_i})\hat{\sigma}_-^i$, where the positive part of the
field operator, $\hat{\mb E}^+(\mb{r})$, at $\mb{r}$, is proportional
to the input field, $\hat{\mb E}^+_\textrm{Input}(\mb{r})$, plus the
field sourced by the atomic operators $\hat{\sigma}_-^i$. Here
$\mb {r_i}$ is the position of atom $i$, and $f(\mb {r},\mb {r_i})$ is
a complex function that depends on the modes of the
system \cite{Solano2017, chang_cavity_2012}.

The probability of measuring two photons at the same time at position
$z$ is proportional to the second-order correlation function \cite{Glauber1963}
\begin{equation}
  \label{eq:g2ssnormalized}
  g^{(2)}(z)=\frac{\langle {\hat{\mb E}}^{(+)}(z)\hat {\mb E}^{(+)}(z){\hat{\mb
    E}}^{(-)}(z){(\hat{\mb
    E}}^{(-)}(z)\rangle}{\textrm{Max}_{\gamma\tau}\left\{\int dz \, \langle {\hat {\mb E}}^{(+)}(z){\hat {\mb
    E}}^{(-)}(z)\rangle  \right\}^2}\, ,
\end{equation}
where we take the maximum over the distance between the atoms, in the
normalization constant, to compare the second-order correlation function for different time lags. By substituting the
input-output relation for the Markovian case into $g^{(2)}$ and
replacing it with the stationary solution, Eq. (\ref{eq:boundstate}),
we see that the dynamics does not lead to a two-photon bound state.
However, the system under consideration is non-Markovian, and one must
calculate the field observables from the full expression for the field
operator.

Figure ~\ref{fig:g2ss} shows the second-order correlation function in
the stationary regime ($t\to\infty$) (see Appendix~\Ref{appcoherence}
for details), for different distances between the atoms. It can be
seen that two photons are trapped between the atoms, a result that is
distinctly non-Markovian, demonstrating the rich interplay of
non-Markovianity and nonlinear atom-photon interactions. Although the
maximum probability of having the atoms excited is given at a distance
$\gamma \tau \approx 0.895$, the probability of having two photons
excited at a particular $z$ is greater at $\gamma \tau \approx 0.375$,
where the instantaneous decay is maximum. This is explained by the
fact that the probability of having two photons trapped in a
particular position increases when the probability of the atomic
excitation diminishes, because the energy from the atomic excitation
should go to the field.

\begin{figure}[h!]
\centering
\includegraphics[width=0.45\textwidth]{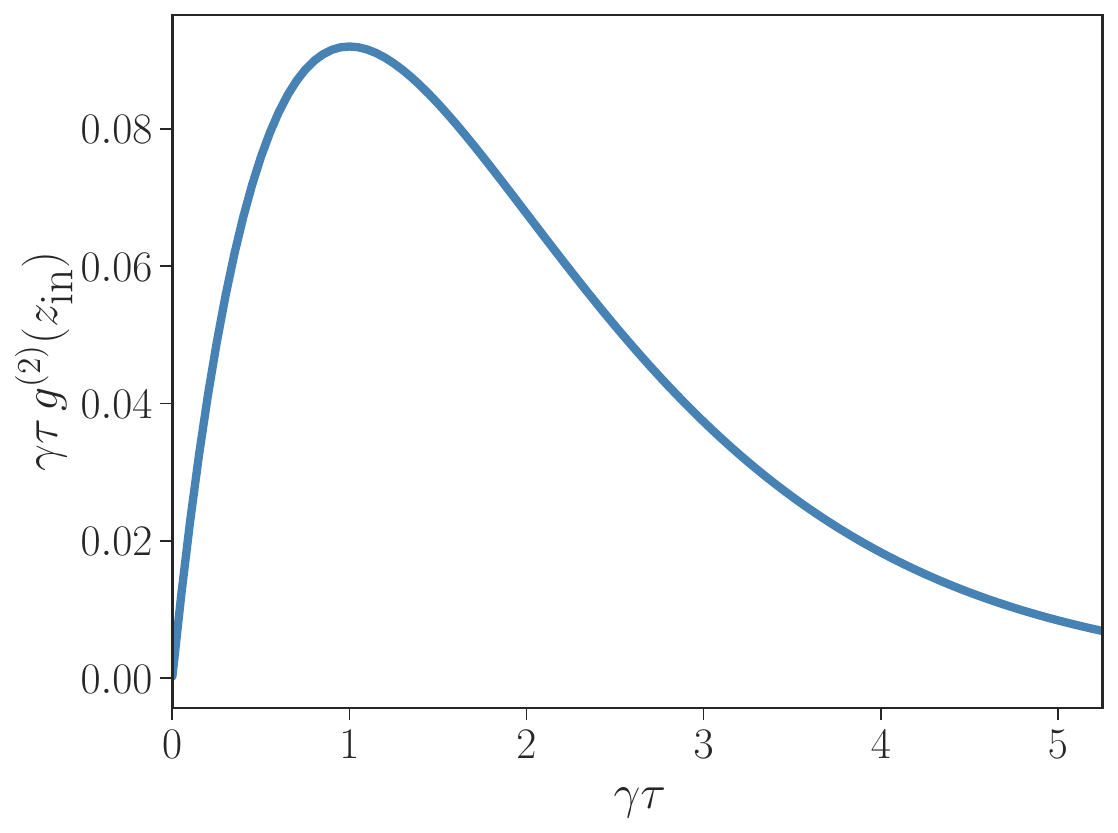}
\caption{\label{fig:Pssin} Probability of detecting two photons in the
  same position, between the atoms, as a function of the delay, for the
  steady state case $\phi=m\pi$.}
\end{figure}

A salient question pertains to the maximum probability of having two
photons between the atoms. This probability depends on the value of
the autocorrelation function and the distance between the two atoms.
Figure~\ref{fig:Pssin} depicts $ \gamma\tau \, g^{(2)}(z_\textrm{in})$,
which is proportional to the probability of having two photons in the
waveguide region separating the two atoms as a function of the delay.
  The distance between the atoms where the probability of having two
  photons is maximum is $\gamma \tau \approx 1.0$.

Note that we chose a normalization that does not depend on the
transverse profile (see Appendix \ref{appcoherence} for details). To
obtain the actual probability that two photon detection happen at the
same time and position, it is useful to calculate the field energy in
the guided field modes, in the limit $t\to\infty$ (see Appendix
\Ref{appstationary})
\begin{eqnarray}
  \label{eq:intensity}
  \mathcal{E}_\textrm{tot} &=& v \epsilon_0  \int d^3\bm{r}\avg{\hat {\mb E}^{(+)}(\bm{r}){\hat {\mb
    E}}^{(-)}(\bm{r})}
\nonumber \\    
&=& \hbar\omega_0(2-P^{(1)}\bkt{t\to\infty})\, , 
\end{eqnarray}
 where $P^{(1)}\bkt{t\to\infty}$ is the probability of having one atomic excitation in the long time limit (see Eq.~(\ref{p1ss})).  This result states that the total field energy corresponds to the energy of two photons with frequency $\omega_0$ (the total energy) minus the energy accumulated in the atoms that has not been emitted into the waveguide.
The normalization constant, $\int dz\langle \hat {\mb E}^{(+)}(z)\hat {\mb E}^{(-)}(z)\rangle$, is given by the total energy around the fiber, $\mathcal{E}_\textrm{total}$ (see Eq.~\eqref{eq:intensity}) multiplied by the square of the mode profile at the atomic frequency transition $\omega_0$.

\begin{figure}[t]
\centering
\includegraphics[width=0.45\textwidth]{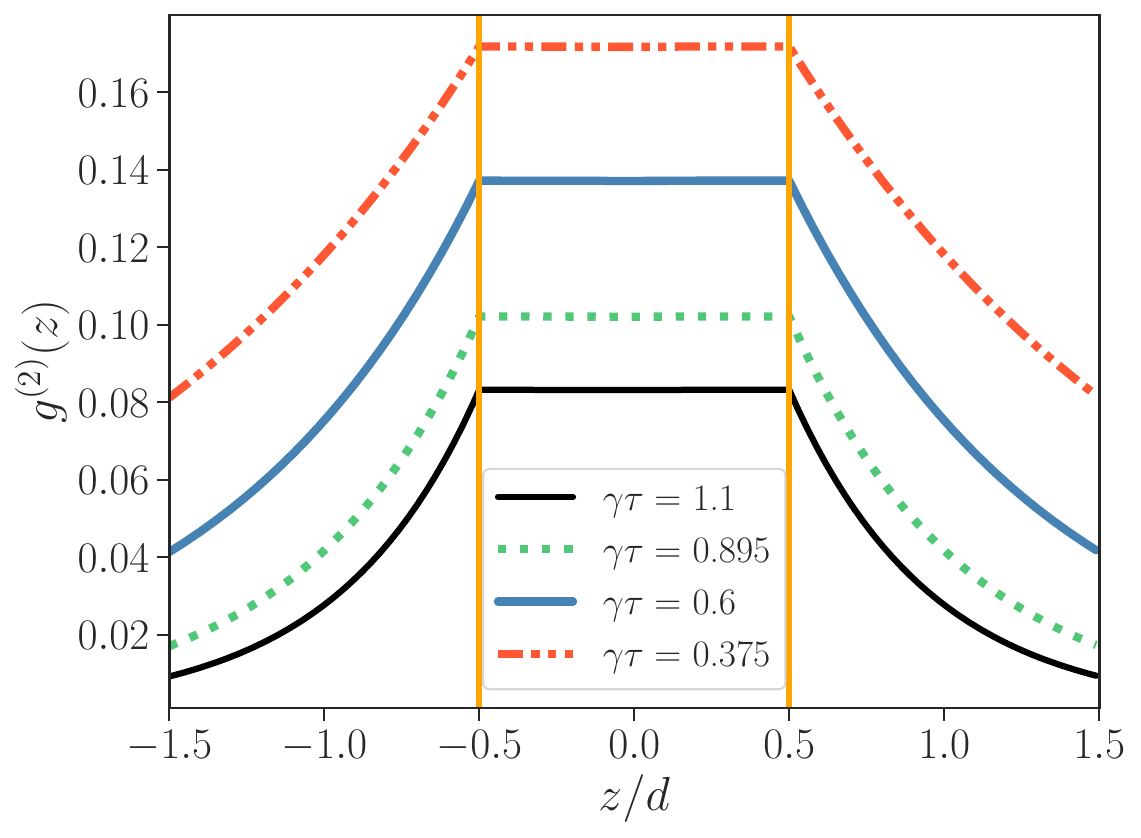}
\caption{\label{fig:g2ss} Second order correlation function as a
  function of the distance between the atoms. The vertical lines serve to show the location of the atoms. }
\end{figure}

\section{Summary and Outlook}
We have analyzed the spontaneous decay of two fully inverted atoms coupled through a waveguide in the presence of retardation effects. A remarkable result is the spontaneous creation of a steady delocalized atom-photon bound state, with sudden birth of entanglement  between the atoms. Furthermore, we demonstrate that such a delay can create two-photon bound states, wherein one can have two photons in the waveguide region between the two atoms. Such  states appear as a result of the non-Markovian time-delayed feedback of the spontaneously radiated EM field acting on the emitters.
Additionally, the collective decay of the two atoms can be momentarily enhanced beyond standard superfluorescence and subsequently inhibited, demonstrating the rich non-Markovian dynamics of such a system.

Such delay-induced spontaneous steady state entanglement generation can have implications in the rapidly growing field of waveguide QED, a field with promising applications in quantum information processing  \cite{Zheng13-2, quantum-network-3, Javadi15, Tian21, Yan15, Coles16, Chang20, Zhong21} that benefit from preparing and manipulating long-lived dark-states. 
In this context, there have been several proposals to generate a steady entangled state; compared to delay-induced \textit{subfluorescence}, these schemes necessitate extra degrees of control, such as external driving fields \cite{Paulisch16, Zanner22,pumping-1, Zheng13}, initial entanglement \cite{sinha}, or chiral emission in front of a mirror \cite{Zhang20, chiral-2}.

It is not only the generation \cite{entanglement-generation} but also the stabilization of entangled states that are  critical to developing efficient quantum devices. In contrast to the idea that the interaction of quantum systems with their environment leads to decoherence and can degrade the entanglement between the components of a quantum system~\cite{decoherence-1}, the environment can also be proposed as a generator \cite{entanglement-by-dissipation-1,   entanglement-by-dissipation-2} and stabilizer of entanglement \cite{entanglement-by-dissipation-3}. Our results demonstrate that non-Markovian time-delayed feedback   can be a mechanism for environment-assisted entanglement generation and stabilization.

Adding delay to the most straightforward collective system of two
two-level atoms leads to different phenomenology, breaking the Born
and Markov approximations, non-trivially modifying the dynamics, and spontaneously creating steady-state quantum correlations between the atoms and multiphoton bound states. Our results further the understanding of the significant role delay plays in quantum optics and present the outset of studying more complex phenomena involving many-body interactions \cite{PhysRevA.68.023809,masson2022,Masson2020}. However, studying this scenario is challenging as the complexity of the problem increases with the number of emitters, and known methods based on master equation approaches fail because neither the Markov nor the Born approximations are valid. Furthermore, although we neglect dispersive dipole-dipole interactions and coupling to non-guided modes to highlight the consequences of delay-induced non-Markovianity, all these effects can coexist, creating richer dynamics that will increase in complexity as the system scales up.

\section{Acknowledgments} We acknowledge helpful discussions with Saikat Guha and Elizabeth Goldschmidt. P.S. is a CIFAR Azrieli Global Scholar in the Quantum Information Science Program. This work was supported by DGAPA-UNAM under grant IG101421 and IG101324 from Mexico, as well as CONICYT-PAI grant 77190033, and FONDECYT grant N$^{\circ}$ 11200192 from Chile. K.S. acknowledges funding support from the John Templeton Foundation Award No. 62422, the National Science Foundation under Grant No. PHY-2309341
and the Army Research Office through Grant No. W911NF2410080.
\vspace{-1 cm}

\appendix
\begin{widetext}
    
\section{Equations of motion: Derivation}

\label{app1}

Using Schr\"odinger equation with the Hamiltonian \eqref{ham} and the state ans\"atz \eqref{eq:psi}, we get the following differential equations for the probability amplitudes
\eqn{
\dot{a}(t) = &- \sum_{r=1}^2 \sum_{\alpha,\eta}   g_a 
\ee{-i\Delta_\alpha t}  b_{\alpha\eta} ^{(r)}(t) \ee{-ik_{\alpha\eta} z_r} \, ,
\label{a-1}}
\eqn{
\dot{b}_{\alpha\eta}^{(r)}(t) = &g_\alpha^* \ee{i\Delta_\alpha t} a(t)
\ee{ik_{\alpha\eta} z_r} 
- \sum_{\beta,\eta '} g_\beta \ee{-i\Delta_\beta t} c_{\alpha\eta, \beta\eta '}(t) \ee{ik_{\beta\eta'} z_r} -\sqrt{2} g_\alpha \ee{-i\Delta_\alpha t} c_{\alpha\eta}(t) \ee{ik_{\alpha\eta} z_r} \, \label{b-1} ,}
\eqn{
\dot{c}_{\alpha\eta, \beta\eta '}(t) =& 2 g_\alpha ^* \ee{i\Delta_\alpha t} \sum_{r=1}^2 b_{\beta\eta '} ^{(r)}(t) \ee{-ik_{\alpha\eta} z_r} + 2 g_\beta ^* \ee{i\Delta_\beta t} \sum_{r=1}^2 b_{\alpha\eta} ^{(r)}(t) \ee{-ik_{\beta \eta '} z_r} \label{c2-1}
\, ,\\
\label{c1-1}
\dot{c}_{\alpha\eta} (t) =& \sqrt{2} g_\alpha ^* \ee{i\Delta_\alpha t} \sum_{r=1}^2 b_{\alpha \eta} ^{(r)}(t) \ee{-ik_{\alpha \eta} z_r} \, ,
}
where we consider atomic positions such that $z_1 = -z_2$.

We integrate \eqref{c2-1} with $c_{\alpha\eta, \beta \eta '} (0) = 0$, and
substitute in \eqref{b-1}, obtaining

\begin{eqnarray*}
\small
\dot{b}_{\alpha \eta}^{(r)}(t) &=& g_\alpha^* \ee{i\Delta_\alpha t} a(t)
\ee{ik_{\alpha\eta} z_r} 
- 2 g_\alpha ^* \sum_{s=1}^2 \int_0 ^t dT \ee{i\Delta_\alpha (t-T)} \ee{-ik_{\alpha \eta} z_s} \sum_{\beta, \eta '} g_\beta \ee{-i\Delta_\beta t} b^{(s)}_{\beta \eta '}(t-T) \ee{ik_{\beta \eta '} z_r} \\
&& 
-\sqrt{2} g_\alpha \ee{-i\Delta_\alpha t}
c_{\alpha\eta}(t) \ee{ik_{\alpha\eta} z_r}
- 
2 \sum_{s=1}^2 \int_0 ^t dT \, b_{\alpha \eta}^{(s)}(t-T) \sum_{\beta, \eta '}  |g_\beta |^2 \ee{-i\Delta_\beta T}  \ee{ik_{\beta \eta'} (z_r-z_s)} \, .
\end{eqnarray*}

For $t\neq 0$, the second term of the RHS can be approximated as zero assuming $g_\beta$ constant and $b_{\beta \eta '} (t)$ evolving slowly, which is consistent with the Wigner-Weisskopf approximation. We transform the sum over frequencies to integrals by using the densities of modes $\rho(w_\alpha)$ \cite{DensityOS}. For guided modes $k_{\alpha \eta} = \eta \frac{\omega_\alpha}{v}$ with $v$ being the phase velocity of the field inside the waveguide, and $\eta = \pm 1$ labels forward or backward propagation direction of the field along the waveguide. Thus $\sum_{\alpha, \eta} \to \sum_{\eta = \pm 1}\int_0^{\infty} d\omega_\alpha \, \rho(\omega_\alpha)$. To study the non-Markovian effects due only to delay and not
to a structured reservoir, we assume a flat spectral density of field
modes around the resonance of the emitters such that
$\omega_\alpha \approx \omega_0$ in the evaluation of $\rho(\omega_\alpha)$ and
$g_\alpha$ functions.

Taking into account all the considerations above we obtain
\begin{eqnarray*}
\sum_{\beta, \eta '} |g_\beta |^2 \ee{-i\omega_\beta T}  \ee{ik_{\beta \eta '} (z_r-z_s)} &=& \int_{0}^{\infty} d\omega_\beta\, \rho(\omega_\beta) |g_\beta|^2 \bigg\{
\ee{-i \omega_\beta (T - \tau_{rs})}
+ \ee{i \omega_\beta (T+\tau_{rs})}
\bigg\} 
\nonumber \\
&\approx& 
\rho(\omega_0) |g_0|^2 \int_{0}^{\infty} d\omega_\beta \bigg\{
\ee{-i \omega_\beta (T - \tau_{rs})}
+ \ee{i \omega_\beta (T+\tau_{rs})}
\bigg\}  \, ,
\end{eqnarray*}

where $\tau_{rs} = \tfrac{z_r - z_s}{v}$. We make use of the Sokhotski-Plemelj theorem
\begin{eqnarray}
\int_0^\infty d\omega_\beta \, \ee{-i\omega_\beta a} = - i \text{PV}\left(\frac{1}{a}\right) + \pi \delta (a) \, ,
\end{eqnarray}
where $\text{PV}$ refers to the Cauchy principal value. Absorbing the contribution of the principal value (which corresponds
to the Lamb shifts) into the atomic transition frequency \cite{superradianceessay} we
obtain
\begin{eqnarray}
\sum_{\beta , \eta '}  |g_\beta |^2 \ee{-i\Delta_\beta T}  \ee{ik_{\beta \eta} (z_r-z_s)} &=& \frac{\gamma \ee{i\omega_0 T}}{2} \frac{\delta(T - \tau_{rs})
 + \delta(T + \tau_{rs})}{2} \, ,
\label{gammas}
\end{eqnarray}
where the single atom decay rate to guided modes is defined as
$\gamma \equiv 4\pi \rho(\omega_0) |g_0|^2$.

Using \eqref{gammas} we obtain the equation of motion for $ \dot{b}_{\alpha \eta}^{(r)}(t)$

\begin{eqnarray*}
\small
\dot{b}_{\alpha \eta}^{(r)}(t) &=& g_\alpha^* \ee{i\Delta_\alpha t} a(t)
\ee{ik_{\alpha \eta} z_r} 
-\sqrt{2} g_\alpha \ee{-i\Delta_\alpha t}
c_{\alpha \eta}(t) \ee{ik_{\alpha \eta} z_r} 
 -\frac{\gamma}{2} b_{\alpha \eta}^{(r)} (t)
- \frac{\gamma}{2} \ee{i\phi} b_{\alpha \eta}^{(s)} (t - \tau) \Theta(t - \tau)\, ,
\qquad r\neq s \, ,
\end{eqnarray*}
with $\tau = |\tau_{12}| = \frac{|z_1-z_2|}{v}$, $\phi=\omega_0 \tau$ and
$\Theta$ is the Heaviside step function.

\section{Equations of motion: Solution}
\label{appsolution}
We use the Laplace transform to solve the delayed differential
equations of motion \eqref{a-2}-\eqref{c1-2}. As an intermediate step, we
define the variables
\begin{eqnarray}
\tilde{B}_{\alpha\eta } ^{(\pm)}(s) = \tilde{B}_{\alpha\eta}^{(1)}(s)\ee{-ik_{\alpha\eta} z_1} \pm \tilde{B}_{\alpha\eta}^{(2)}(s)\ee{-ik_{\alpha\eta} z_2} \,.
\end{eqnarray}
Thus
\begin{align}
s\, \tilde{a}(s) - 1 &= -\sum_{\alpha, \eta} g_\alpha \tilde{B}_{\alpha \eta}^{(+)}(s)
\label{a12-ec}
\\
\left[s + i\Delta_\alpha +\frac{\gamma}{2} + \frac{4|g_\alpha |^2}{s+2i\Delta_\alpha} \right] \tilde{B}_{\alpha\eta}^{(+)}(s) &= 2g_\alpha^{*} \tilde{a}(s)-\frac{\gamma}{2}
 \ee{i\phi}\ee{-i\Delta_\alpha \tau - s\tau}\left\{ 
\cos(k_{\alpha \eta} d) \tilde{B}_{\alpha\eta}^{(+)}(s) + i \sin(k_{\alpha \eta} d) \tilde{B}_{\alpha \eta}^{(-)}(s) 
\right\}
\label{bp-ec}
\\
\left[s + i\Delta_\alpha +\frac{\gamma}{2} \right] \tilde{B}_{\alpha\eta}^{(-)}(s) &=
\frac{\gamma}{2} \ee{i\phi}\ee{-i\Delta_\alpha \tau - s\tau}
\left\{ 
\cos(k_{\alpha\eta} d) \tilde{B}_{\alpha\eta}^{(-)}(s) + i \sin(k_{\alpha \eta} d) \tilde{B}_{\alpha \eta}^{(+)}(s) 
\right\} \, ,
\label{bm-ec}
\end{align}
where we use that
\begin{eqnarray*}
\tilde{C}_{\alpha\eta} (s) = \sqrt{2} g_\alpha^{*} \frac{\tilde{B}_{\alpha\eta}^{(+)}(s)}{s+2i\Delta_\alpha} \,,
\end{eqnarray*}
with $a(0)=1$, $B_{\alpha\eta}^{(1,2)}(0) = 0$ and $C_{\alpha\eta}(0) = 0$ as initial conditions.

The solutions of the previous equations are

\eqn{
\label{bp-sol}
\tilde{B}_{\alpha\eta}^{(+)}(s) & 
= - 2g_\alpha^{*} \tilde{a}(s)
\sbkt{s + i\Delta_\alpha + \frac{\gamma}{2} - \frac{\gamma}{2}e^{i\phi} e^{ -(s+ i \Delta_\alpha) \tau } \cos(k_{\alpha\eta} d)}\sbkt{\cbkt{ 
\frac{\gamma}{2}e^{i\phi} \ee{ -(s + i \Delta_\alpha) \tau }
\sin(k_{\alpha \eta} d) }^2 \right.\non\\
&\left.+\cbkt{ s + i\Delta_\alpha + \frac{\gamma}{2} -\frac{\gamma}{2}e^{i\phi} e^{ - (s+i \Delta_\alpha) \tau } \cos(k_{\alpha\eta} d)} \cbkt{ s + i\Delta_\alpha + \frac{\gamma}{2} +\frac{\gamma}{2}e^{i\phi} e^{ -(s + i \Delta_\alpha ) \tau} \cos(k_{\alpha\eta} d) + \frac{4 \lvert g_\alpha \rvert ^2}{s+2i\Delta_\alpha}
} }^{-1} 
}
\eqn{
\label{bm-sol}
\tilde{B}_{\alpha\eta}^{(-)}(s) &= - 2g_\alpha^{*} \tilde{a}(s)
\sbkt{i \frac{\gamma}{2}e^{i\phi} e^{ -(s+ i \Delta_\alpha) \tau } \sin(k_{\alpha \eta} d)}\sbkt{\cbkt{ 
\frac{\gamma}{2}e^{i\phi} \ee{ -(s + i \Delta_\alpha) \tau }
\sin(k_{\alpha \eta} d)}^2\right. \non\\
&\left.+ \cbkt{ s + i\Delta_\alpha + \frac{\gamma}{2} -\frac{\gamma}{2}e^{i\phi} e^{ - (s+i \Delta_\alpha) \tau } \cos(k_{\alpha\eta} d)} \cbkt{ s + i\Delta_\alpha + \frac{\gamma}{2} +\frac{\gamma}{2}e^{i\phi} e^{ -(s + i \Delta_\alpha ) \tau} \cos(k_{\alpha\eta} d) + \frac{4 \lvert g_\alpha \rvert ^2}{s+2i\Delta_\alpha}
}  } \,.
}
and 
\eqn{
\label{a12-sol}
\tilde{a}(s) &= \sbkt{ s - 2 \sum_{\alpha, \eta} |g_\alpha |^2 
 \sbkt{s + i\Delta_\alpha + \frac{\gamma}{2} - \frac{\gamma}{2}e^{i\phi} e^{-(s + i \Delta_\alpha) \tau } \cos(k_{\alpha \eta} d)}\sbkt{\cbkt{\frac{\gamma}{2}e^{i\phi} e^{- (s + i \Delta_\alpha) \tau }\sin(k_\alpha d) }^2 \right.\right.\non\\
 &\left.\left.+\cbkt{ s + i\Delta_\alpha + \frac{\gamma}{2} -\frac{\gamma}{2}e^{i\phi} e^{-(s + i \Delta_\alpha) \tau }\cos(k_{\alpha \eta} d)} \cbkt{ s + i\Delta_\alpha + \frac{\gamma}{2} +
\frac{\gamma}{2}e^{i\phi} e^{-(s + \Delta_\alpha) \tau }
\cos(k_{\alpha \eta} d) + \frac{4 \lvert g_\alpha \rvert ^2}{s+2i\Delta_\alpha}
}  } }^{-1} \,.
} 
In order to find the inverse Laplace transform of Eqs. \eqref{bp-sol}, \eqref{bm-sol} and \eqref{a12-sol} we use $ |g_\alpha|^2 / \gamma \sim 10^{-4}$ for EM modes in the visible range \cite{Barnes2020} to neglect the contribution of the term $4|g_\alpha|^2 /(s + i \Delta_\alpha)$ in the denominator. This corresponds to neglecting the probability of exciting two equal $\alpha$ modes traveling in the same direction compared to the probability of exciting two different modes. Considering this, we get Eq. \eqref{a12-solapp} and

\eqn{\label{bs}
\tilde{B}_\alpha^{\bkt{\substack{1\\2}}}(s) &\approx& - g_\alpha^{*} \tilde{a}(s) e^{\pm i k_{\alpha \eta} d/2}
  \frac{s + i\Delta_\alpha + \frac{\gamma}{2} - \frac{\gamma}{2}e^{i\phi} e^{-s \tau }e^{ - i \Delta_\alpha \tau} \ee{\mp ik_{\alpha\eta} d}}{\left[ 
\frac{\gamma}{2}e^{i\phi} e^{-s \tau }e^{ - i \Delta_\alpha \tau} + s + i\Delta_\alpha + \frac{\gamma}{2}
\right] \left[ 
\frac{\gamma}{2}e^{i\phi} e^{-s \tau }e^{ - i \Delta_\alpha \tau} - s - i\Delta_\alpha - \frac{\gamma}{2}
\right] } \, .
}

\subsection{Solving the integrals for $\tilde{a}(s)$}\label{app:sol-a}

Equation \eqref{a12-solapp} with the sum approximated as an integral is 
\begin{eqnarray}
\label{a12int}
\tilde{a}(s) &\simeq & \left\{ s - \frac{\gamma}{\pi} \int_{-\omega_0}^{\infty}
 \frac{s + i\Delta_\alpha + \frac{\gamma}{2} - \frac{\gamma}{2}e^{i\phi} e^{-s \tau }e^{ - i \Delta_\alpha \tau} \cos(\Delta_\alpha \tau + \phi)}{\left[ 
\frac{\gamma}{2}e^{i\phi} e^{-s \tau }e^{ - i \Delta_\alpha \tau} + s + i\Delta_\alpha + \frac{\gamma}{2}
\right] \left[ 
\frac{\gamma}{2}e^{i\phi} e^{-s \tau }e^{ - i \Delta_\alpha \tau} - s - i\Delta_\alpha - \frac{\gamma}{2}
\right] } d\Delta_\alpha \right\}^{-1} \, 
\nonumber \\
&=& \left\{ s - \frac{\gamma}{\pi} I(s, \gamma, \tau, \phi)
\right\}^{-1} \, ,
\end{eqnarray}
where we extend the lower limit in the integral to $-\infty$ as an approximation. We
rewrite the denominator inside the integral using the poles for the variable $\Delta_\alpha$,
determined by the characteristic equation,
\begin{eqnarray}
\frac{\gamma}{2}e^{i\phi} e^{-s \tau }e^{ - i \Delta_\alpha \tau} -\sigma \left( s + i\Delta_\alpha + \frac{\gamma}{2}\right) = 0
\qquad \Longrightarrow \qquad \Delta^{(\sigma)}_k = i \left\{s + \frac{\gamma}{2} - \frac{1}{\tau} W_k(\sigma r) \right\} \, ,
\end{eqnarray}
with $W_k$ denoting the $k$th branch of Lambert W function
\cite{Corless1996} and $\sigma = \pm 1$. For simplicity, we introduce
$r=\frac{\gamma \tau}{2}\ee{\gamma \tau /2 + i\phi}$. Using partial fraction
decomposition \cite{partial-fd} we obtain
\eqn{
\label{pfd}
  \frac{1}{\left[ 
\frac{\gamma}{2}e^{i\phi} e^{-s \tau }e^{ - i \Delta_\alpha \tau} + s + i\Delta_\alpha + \frac{\gamma}{2}
\right]\left[ 
\frac{\gamma}{2}e^{i\phi} e^{-s \tau }e^{ - i \Delta_\alpha \tau} - s - i\Delta_\alpha - \frac{\gamma}{2}
\right] } 
= \frac{\tau}{2} \sum_{k=-\infty}^{\infty} \sum_{\sigma =\pm 1}
\frac{1}{W_k(\sigma r) + W_k^{2}(\sigma r)}\frac{i}{\Delta_\alpha - \Delta_{k}^{(\sigma)}} \, .
}

Using Cauchy's integral formula we get
\begin{eqnarray}
\int_{-\infty}^{\infty} \frac{s + i\Delta_\alpha + \frac{\gamma}{2}}{\Delta_\alpha - \Delta_k^{(\sigma)}} d \Delta_\alpha = 2\pi i \frac{W_k{(\sigma r)}}{\tau}  \, ,
\\
\int_{-\infty}^{\infty} \frac{e^{-i\Delta_\alpha \tau} \cos(\Delta_\alpha \tau + \phi)}{\Delta_\alpha - \Delta_k^{(\sigma)}} d \Delta_\alpha = 2\pi i \ee{i\phi} \, .
\end{eqnarray}
For the second integral we use that
$\textrm{Im}\left(\Delta_k ^{(\sigma)} \right) > 0$ because we can
take $\textrm{Re}(s)$ as large as we need, obtaining {\small
\begin{flalign}
\label{int-i}
I(s, \gamma, \tau, \phi) = -\pi \sum_{k=-\infty}^{\infty} \sum_{\sigma = \pm 1}
\frac{1}{1+ W_k(\sigma r)}\left(1 + \frac{\tau \ee{i\phi}}{2 W_k(\sigma r)}  \right) \,.
\end{flalign} }
Using the following identities of the Lambert functions\cite{Lambert-identities}
\begin{eqnarray}
\sum_{k=-\infty}^{\infty} \frac{1}{1+W_k(z)} &=& \frac{1}{2} \, ,
\\
\sum_{k=-\infty}^{\infty} \frac{1}{W_k(z) + W_k^2(z)}
&=& \frac{1}{z} \,,
\end{eqnarray}
we obtain $I(s, \gamma, \tau, \phi) = - \pi$.

\subsection{Inverse Laplace transform of $\tilde{B}_{\alpha\eta}^{(1,2)}$} 

To obtain $B_{\alpha \eta}^{(1,2)}(t)$ we apply the inverse Laplace transform of Eq.~\eqref{bs}. First, similar to what is done in Appendix \ref{app:sol-a}, we rewrite the result using the poles of the denominator but for the variable $s$. 
Defining
\begin{eqnarray}
\label{poles-s}
s_{k,\alpha}^{(\sigma)} = - \frac{\gamma_{k\sigma}}{2} - i\Delta_\alpha \qquad 
\textrm{with} \qquad \gamma_{k\sigma} = \gamma  - \frac{2}{\tau} W_k(\sigma r)\, ,
\end{eqnarray}
and using that $k_{\alpha \eta} d = \eta (\Delta_\alpha \tau +
\phi)$ and $\tilde{a}(s)=1/(s+\gamma)$, we obtain
{\small
\begin{eqnarray}
B_{\alpha \eta}^{\bkt{\substack{1\\2}}}(t) = -\frac{g_\alpha^{*}\tau}{2}
\sum_{\sigma = \pm 1} \sum_{k=\infty}^{\infty}  
\frac{\ee{\pm ik_{\alpha \eta} d/2}}{W_k(\sigma r) + W_k^{2}(\sigma r)}
\bigg[
\mathcal{L}^{-1} \left(
\frac{s + i\Delta_\alpha + \frac{\gamma}{2}}{(s+\gamma)(s-s_{k,\alpha}^{(\sigma)})} \right)
- 
\frac{\gamma}{2} \ee{i\phi \mp k_{\alpha\eta} d}
\mathcal{L}^{-1} \left(
\frac{e^{-s\tau}}{(s+\gamma)(s-s_{k,\alpha}^{(\sigma)})} \right)
\bigg]
\end{eqnarray}}
where $\mathcal{L}^{-1}$ denotes the inverse Laplace transform. Using that
\begin{eqnarray}
\mathcal{L}^{-1} \left(
\frac{s + i\Delta_\alpha + \frac{\gamma}{2}}{(s+\gamma)(s-s_{k, \alpha}^{(\sigma)})} \right)
&=& \frac{\frac{\gamma}{2} - i \Delta_\alpha }{s_{k,\alpha}^{(\sigma)}+\gamma} \ee{-\gamma t} + 
\frac{\frac{\gamma}{2} + i \Delta_\alpha +s_{k,\alpha}^{(\sigma)} }{s_{k,\alpha}^{(\sigma)}+\gamma} \ee{s_{k,\alpha}^{(\sigma)} t}
\\
\mathcal{L}^{-1} \left(
\frac{\ee{-s\tau}}{(s+\gamma)(s-s_{k,\alpha}^{(\sigma)})} \right)
&=&
\left( \ee{s_{k,\alpha}^{(\sigma)} (t-\tau)} -
\ee{-\gamma(t-\tau)} \right) \frac{\Theta(t-\tau) }{s_{k,\alpha}^{(\sigma)}+\gamma}
\end{eqnarray}
and taking $b_{\alpha\eta}^{(r)} = B_{\alpha\eta} ^{(r)} (t) \ee{i\Delta_\alpha
  t}$ we get
\begin{eqnarray}
\label{bt}
b_{\alpha \eta}^{\bkt{\substack{1\\2}}}(t) &=& - \frac{g_\alpha^{*} \tau}{2} \ee{i\Delta_\alpha t} e^{\pm i \eta (\Delta_\alpha \tau + \phi)/2} \sum_{\sigma = \pm 1} \sum_{k=-\infty}^{\infty}  \frac{1}{W_k(\sigma r) + W_k^2(\sigma r)}
\frac{\ee{-\gamma t}}{\gamma - \frac{\gamma_{k\sigma}}{2} - i\Delta_\alpha}
\bigg[
\frac{\gamma}{2}-i\Delta_\alpha
\nonumber
 \\
&&
 + 
\frac{W_k(\sigma r)}{\tau}\ee{\left[\gamma - \gamma_{k\sigma}/2 - i \Delta_\alpha \right] t}
+ \frac{\gamma}{2} \ee{i(1-\eta)\phi - i(1+\eta) \Delta_\alpha \tau + \gamma\tau}
\Theta(t-\tau) \bigg\{1-\ee{\left[\gamma - \gamma_{k\sigma}/2 - i \Delta_\alpha \right] (t-\tau)}  \bigg\}
\bigg]  \, ,
\end{eqnarray}
where $W_k$ represents the $k$-th branch of the Lambert-W function.

\subsection{Solution for $c_{\alpha \eta, \beta \eta '} (t)$}
Applying the Laplace transform to equation \eqref{c2-1} and defining $C_{\alpha \eta, \beta \eta '}(t) = c_{\alpha \eta, \beta \eta '}(t)
\ee{-i(\Delta_\alpha + \Delta_\beta)t}$ we get
\begin{eqnarray}
\tilde{C}_{\alpha\eta, \beta\eta '}(s)
&=& 2 g_\alpha ^* \sum_{r=1}^2 \frac{\tilde{B}_{\beta\eta '} ^{(r)}(s) \ee{-ik_{\alpha \eta } z_r}}{s+i(\Delta_\alpha + \Delta_\beta)} + 2 g_\beta ^*  \sum_{r=1}^2 \frac{ \tilde{B}_{\alpha\eta} ^{(r)}(s) \ee{-ik_{\beta \eta '} z_r}}{s+i(\Delta_\alpha + \Delta_\beta)} \, .
\end{eqnarray}
To find the function in the time domain,
we substitute Eq.~\eqref{bs} and use the poles \eqref{poles-s}. Then, applying the inverse Laplace transform, we arrive at the expression
\begin{eqnarray}
\label{eq:c-solution}
c_{\alpha \eta, \beta \eta '}(t) 
&=&
2\tau g_\alpha^{*}g_\beta^{*} \sum_{\sigma = \pm 1} \sum_{k=-\infty}^{\infty}
\frac{\ee{i(\Delta_\alpha + \Delta_\beta)t}}{W_k(\sigma r) + W_k^{2}(\sigma r)}
\bigg[
\cos \left[(k_{\alpha \eta}- k_{\beta \eta '})d/2 \right]
\left\{ M_{\alpha, \beta}^{(k,\sigma)}(t) + M_{\beta, \alpha}^{(k,\sigma)}(t)  \right\}
\nonumber \\
&& - \frac{\gamma}{2} e^{i\phi}  \cos \left[(k_{\alpha \eta} + k_{\beta \eta '})d/2 \right]
\left\{N_{\alpha, \beta}^{(k,\sigma)}(t) \ee{-i\Delta_\alpha \tau}+ N_{\beta, \alpha}^{(k,\sigma)}(t) \ee{-i\Delta_\beta \tau} \right\}
\bigg] \label{eq:calpha}\, ,
\end{eqnarray}
where we have defined the functions
\begin{eqnarray}
M_{\alpha, \beta}^{(k,\sigma)}(t) &=& 
\frac{\gamma - 2 i \Delta_\alpha}{\left[\Delta_\alpha + i \left( \gamma -
\frac{\gamma_{k\sigma}}{2}
\right) \right] \left[\Delta_\alpha + \Delta_\beta +i\gamma \right]} \frac{\ee{-\gamma t}}{2} 
- \frac{\gamma - 2 i \Delta_\beta}{\left[\Delta_\beta + i
\frac{\gamma_{k\sigma}}{2}  \right] \left[\Delta_\alpha + \Delta_\beta +i\gamma \right]} \frac{\ee{-i(\Delta_\alpha + \Delta_\beta) t}}{2}
\nonumber \\
&&  
+
\frac{W_k(\sigma r)}{\tau} \frac{\ee{-\gamma_{k\sigma} t/2 - i \Delta_\alpha t}}{\left[
\Delta_\alpha + i\left(\gamma - \frac{\gamma_{k\sigma}}{2} \right)
\right]\left[
\Delta_\beta + i\frac{\gamma_{k\sigma}}{2} \right]} \, ,
\\
N_{\alpha, \beta}^{(k,\sigma)}(t) &=&- \Bigg(
\frac{\ee{-\gamma (t-\tau)}}{\left[\Delta_\alpha + i \left(\gamma -
\frac{\gamma_{k\sigma}}{2} 
\right) \right] \left[\Delta_\alpha + \Delta_\beta +i\gamma \right]}  
+ \frac{\ee{-i(\Delta_\alpha + \Delta_\beta) (t-\tau)}}{\left[\Delta_\beta + i 
\frac{\gamma_{k\sigma}}{2} \right] \left[\Delta_\alpha + \Delta_\beta +i\gamma \right]} 
\nonumber \\
&&  
-
\frac{\ee{-\gamma_{k\sigma} (t-\tau)/2 - i \Delta_\alpha (t-\tau)}}{\left[
\Delta_\alpha + i\left(\gamma - \frac{\gamma_{k\sigma}}{2} \right)
\right]\left[
\Delta_\beta + i\frac{\gamma_{k\sigma}}{2} 
\right]} \Bigg) \Theta(t-\tau) \,.
\end{eqnarray}
These functions are not invariant to the interchange of labels $\alpha, \beta$.

\section{Stationary values}
\label{appstationary}
In the late-time limit $\gamma t \to \infty$ we have for \eqref{bt} the value
\begin{eqnarray}
b_{\alpha\eta}^{\bkt{\substack{1\\2}}}(t) \approx -\frac{g_\alpha^{*} \tau}{2}  \sum_{\sigma = \pm 1}\sum_{k=\infty}^{\infty} 
\frac{\ee{\pm i\eta (\Delta_\alpha \tau + \phi )/2}}{W_k(\sigma r) + W_k^2(\sigma r)} 
\frac{\ee{-\left[\frac{1}{2} - \frac{1}{\gamma \tau}W_{k}(\sigma r) \right]\gamma t}}{\frac{\gamma}{2} - i \Delta_\alpha + \frac{1}{\tau}W_k(\sigma r)}
\bigg[
\frac{W_k(\sigma r)}{\tau} - 
\frac{\gamma}{2}\ee{i(1-\eta)\phi}
\ee{-i\eta \Delta_\alpha \tau}
\ee{-\left[\frac{1}{2} - \frac{1}{\gamma \tau}W_{k}(\sigma r) \right]\gamma \tau}
 \bigg] \, .
 \nonumber
\end{eqnarray}
The above expression does not decay to zero if
\begin{eqnarray}
\Re \left[\frac{1}{2} - \frac{1}{\gamma \tau} W_k\left(\sigma \frac{\gamma \tau}{2} \ee{\gamma \tau /2} \ee{i\phi} \right) \right] = 0 \, .
\end{eqnarray}
Thus, the terms of the expression that will be non-zero in the long
time limit are $k=0$, and $\{\sigma=+1, \, \phi=2\pi n\}$ or
$\{\sigma=-1, \, \phi=(2n+1)\pi\}$, with $n \in \mathbb{N}^{0}$,
because
\begin{eqnarray*}
W_0\left(\frac{\gamma \tau}{2}\ee{\gamma \tau /2} \right) = \frac{\gamma \tau}{2} \, . 
\end{eqnarray*} 

Taking into account the above, we get a stationary value given by
\begin{eqnarray}
b_{\alpha\eta,\, \text{ss}}^{\bkt{\substack{1\\2}}} = -\frac{g_\alpha^{*}}{2}  \frac{\ee{\pm i\eta (\Delta_\alpha \tau + n \pi )/2}}{1 + \frac{\gamma \tau}{2}} 
\frac{1 - \ee{ \mp i \eta \Delta_\alpha \tau }}{\gamma - i \Delta_\alpha} \, .
\end{eqnarray}

Then, we obtain
\begin{eqnarray}
P^{(1)}\bkt{t\to\infty} &=& \frac{\rho(\omega_0)|g_0|^{2}}{2(1+\frac{\gamma \tau}{2})^{2}}
\sum_{\eta=\pm 1} \int_{-\infty}^{\infty} d\Delta_\alpha \left\lvert
\frac{1 - \ee{ - i \eta\Delta_\alpha \tau}}{\gamma - i \Delta_\alpha}
\right \rvert^{2} \, ,
\nonumber
\end{eqnarray}
which leads to Eq.~\eqref{p1ss}.

For correlations between atomic dipoles, we find the stationary value
\begin{eqnarray}
\tr\sbkt{\hat \rho_A (t\to \infty)\hat{\sigma}_+ ^{(1)} \hat \sigma_-^{(2)} } &=& \frac{\rho(\omega_0)|g_0|^{2} \cos(n\pi)}{4(1+\frac{\gamma \tau}{2})^{2}}
\sum_{\eta=\pm 1} \int_{-\infty}^{\infty} d\Delta_\alpha 
\frac{\left(1 - \ee{ - i \eta\Delta_\alpha \tau} \right)^{2}}{\left\lvert\gamma - i \Delta_\alpha \right \rvert^{2}}
= -\frac{\cos(n\pi)}{2} \frac{\sinh \left(\frac{\gamma\tau}{2} \right)}{(1+\frac{\gamma\tau}{2})^2 \ee{\frac{\gamma\tau}{2}}} \, .
\end{eqnarray}
Therefore $\tr\sbkt{\hat \rho_A (t \to \infty)\hat{\sigma}_+ ^{(1)} \hat \sigma_-^{(2)} } = P^{(1)}(t\to \infty)/2$.

Finally, the function $c_{\alpha\eta, \beta \eta '}(t)$ from \eqref{eq:c-solution} has a simplified expression in the long-time limit $\gamma t \to \infty$ and $\phi = n \pi$. In this scenario, the value is equal to
\begin{eqnarray}
\label{Eq:cinf}
c_{\alpha \eta, \beta \eta '}(t\to \infty) 
&=&
 g_\alpha^{*}g_\beta^{*} 
\sum_{u = \pm 1} \sum_{r = 1}^2
\ee{- i (\eta + u \eta ')\phi \bar{z}_r }
\ee{- i (\eta \Delta_\alpha + u \eta ' \Delta_\beta)\bar{z}_r \tau }
\bigg[
\frac{\bar{A}^{(u)}}{1 + \frac{\gamma \tau}{2}}
\left(
\frac{\ee{i\Delta_\alpha t}}{\Delta_\alpha [\Delta_\beta + i\gamma]}
+
\frac{\ee{i\Delta_\beta t}}{\Delta_\beta [\Delta_\alpha + i\gamma]}
\right)
\nonumber\\
&&
 - \sum_{\sigma = \pm 1} \sum_{k=-\infty}^{\infty} \frac{A_{k\sigma}^{(u)}}{1 + W_k(\sigma r)}
\bigg( 
\frac{\ee{\delta_{u,1} i \Delta_\alpha\tau }}{[\Delta_\alpha + \Delta_\beta + i\gamma ][\Delta_\alpha + i \frac{\gamma_{k,\sigma}}{2} ]}
+
\frac{\ee{\delta_{u,1} i \Delta_\beta\tau }}{[\Delta_\alpha + \Delta_\beta + i\gamma ][\Delta_\beta + i \frac{\gamma_{k,\sigma}}{2}]}
\bigg)
\bigg] 
\end{eqnarray}
where we have the auxiliary value $u= \pm 1$ and the coefficients $\bar{A}^{(-1)} = 1$, $\bar{A}^{(+1)} = - \ee{i\phi}$, $A_{k\sigma}^{(-1)} = 1$ and $A_{k\sigma}^{(+1)} = -\gamma\tau \ee{i\phi} /(2W_k(\sigma r))$. In addition, we introduced the normalized position of atoms $\bar{z}_{1,2} = z_{1,2}/d = \pm 1/2$.

\section{Second-order correlation function}
\label{appcoherence}
We can write the field operator, in cylindrical coordinates, as \cite{Lekienmodes}:
\begin{equation}
  \label{fieldop}
  \hat{\mb E}^{(+)}(z, t)=i\sum_{\alpha, \eta} \mathcal{E}_{\alpha} 
  \hat{\mb e}_{\alpha}(r,\varphi)
  \hat{a}^\dagger_{\alpha, \eta} \ee{-i k_{\alpha \eta}z} 
\end{equation}
with
$\mathcal{E}_{\alpha} = \sqrt{\hbar\omega_\alpha/(4\pi v \epsilon_0)}$
and ${\mb e}_{\alpha}(r, \varphi)$  the transverse profile function,
which satisfies the normalization condition,
\begin{eqnarray}
\label{normalized-e}
\int_0^{2\pi} d\varphi \int_{0}^\infty \left| {\mb e}_{\alpha} \right|^2n^2(r)  r\,dr = 1  \, ,
\end{eqnarray}
where $n^2(r)$ is the refractive index of the cylindrical waveguide.

To calculate the correlation function
$G^{(2)}(z, z) = \langle \hat{\mb E}^{(-)}(z, t) \hat{\mb E}^{(-)}(z,
t) \hat{\mb E}^{(+)}(z, t) \hat{\mb E}^{(+)}(z, t) \rangle$, for the
position $z$ along the waveguide and in the long-time limit, we use the
Wigner-Weisskopf approximation (see Appendix \ref{app1}) to evaluate
$\mathcal{E}_{\alpha} \approx \mathcal{E}_{0}$ and
${\mb e}_{\alpha} \approx {\mb e}_{0}$, and the state
\eqref{eq:boundstate}. We get the following,
\begin{eqnarray*}
G^{(2)}(z, z) = |\mathcal{E}_{0}|^4 |{\mb e}_{0}|^4 \bigg| \sum_{\alpha, \eta}
\sum_{\beta, \eta'} c_{\alpha \eta, \beta \eta '}(t\to \infty)
\ee{-i(k_{\alpha\eta} + k_{\beta \eta'})z}
 \bigg|^2 \, .
\end{eqnarray*}
Introducing Eq.~(\ref{Eq:cinf}) and performing the integrals in $\Delta_\alpha$, $\Delta_\beta$ we obtain
\begin{eqnarray}
\frac{G^{(2)}(z, z)}{\rho_0^2 |\mathcal{E}_{0}|^4 |{\mb e}_{0}|^4 }& = & \pi^2 \gamma^2 \bigg| \sum_{u = \pm 1} \sum_{r = 1}^2
\sum_{\eta, \eta'} 
\sum_{k=-\infty}^{\infty} \frac{A_{k\sigma}^{(u)}}{1 + W_k(\sigma r)}
\bigg[\ee{-\eta [\bar{z} + \bar{z}_r] \gamma \tau }\Theta(\eta [\bar{z} + \bar{z}_r]) \times
\nonumber \\
&&
\ee{-(\bar{z} [\eta' - \eta] + \bar{z}_r [u\eta' - \eta]  - \delta_{u,1} ) \gamma_{k\sigma} \tau/2}
\Theta(\bar{z} [\eta' - \eta] + \bar{z}_r [u\eta' - \eta] - \delta_{u,1}) +
\ee{-\eta' [\bar{z} + u \bar{z}_r] \gamma \tau } 
\Theta(\eta' [\bar{z} + u\bar{z}_r])\times
\nonumber \\ &&
\ee{-(\bar{z} [\eta - \eta '] + \bar{z}_r [\eta - u\eta'] - \delta_{u,1} ) \gamma_{k\sigma} \tau/2}
\Theta(\bar{z} [\eta - \eta '] + \bar{z}_r [\eta - u\eta'] - \delta_{u,1})
\bigg]
 \bigg|^2 \,,
\end{eqnarray}
with $\bar{z} = z/d$.
To normalize this function, we use
\begin{eqnarray*}
\int_{-\infty} ^\infty dz \, \langle { \hat {\mb E}}^{(+)}(z){\hat {\mb  E}}^{(-)}(z) \rangle = |\mathcal{E}_{0}|^2 |{\mb e}_{0}|^2
\int_{-\infty} ^\infty dz
\sum_{\alpha, \eta}
\sum_{\beta, \eta'} 
\ee{-i(k_{\alpha\eta} - k_{\beta\eta'}) z} \langle
\hat{a}^\dagger _{\beta,\eta'}
\hat{a} _{\alpha,\eta}
 \rangle \, .
\end{eqnarray*}
Again, we have considered  $\mathcal{E}_{\alpha} \approx \mathcal{E}_{0}$ and ${\mb e}_{\alpha} \approx {\mb e}_{0}$.  Then, using the identity,
\begin{eqnarray*}
\int_{-\infty} ^\infty dz \, \ee{-i(k_{\alpha\eta} - k_{\beta\eta'})z}
= 2\pi \left[
\delta(\eta - \eta')\delta(\omega_\alpha - \omega_\beta)
+ \delta(\eta + \eta')\delta(\omega_\alpha + \omega_\beta)
\right]\, ,
\end{eqnarray*}
we arrive at the expression,
\begin{eqnarray}
\frac{1}{\rho_0 |\mathcal{E}_{0}|^2 |{\mb e}_{0}|^2}\int_{-\infty} ^\infty dz \, \langle { \hat {\mb E}}^{(+)}(z){\hat {\mb E}}^{(-)}(z) \rangle &=& 
4\pi \sum_{\alpha, \eta}
\bigg( 
\sum_{r=1}^2 \left| b_{\alpha\eta,\, \text{ss}}^{(r)} \right|^2
+ \sum_{\beta, \eta '}
\left| c_{\alpha \eta, \beta \eta '}(t\to \infty) \right|^2 
\bigg)
\nonumber \\
&=& 4\pi (2 - P^{(1)}\bkt{t\to\infty} ) \,
\end{eqnarray}
where we have used 
\begin{eqnarray*}
\sum_{\alpha, \eta}\sum_{\beta, \eta '}
\frac{\left| c_{\alpha \eta, \beta \eta '}(t) \right|^2}{2} = 1 - P^{(2)}(t) - P^{(1)}(t) \, . 
\end{eqnarray*}
\end{widetext}

\bibliography{apssamp}

\end{document}